\documentclass[preprint,authoryear,12pt]{elsarticle}
\usepackage{amssymb}
\usepackage{times}
\usepackage{amsmath}
\usepackage{epsfig}
\usepackage{amsfonts}
\usepackage{verbatim}
\usepackage{amsthm}
\usepackage{graphicx}
\usepackage{subfigure}
\usepackage{booktabs}
\usepackage{helvet}
\usepackage{courier}
\usepackage{color}
\usepackage[linesnumbered,ruled,vlined]{algorithm2e}
\usepackage{array}
\newcommand{\eat}[1]{}

\usepackage{bm}
\usepackage{multirow}
\usepackage{hyperref}
\usepackage{cases}
\usepackage{caption}

\usepackage{algorithmic}
\newcommand{\etal}{{\em et al.~}}       
\newcommand{\eg}{{\em e.g.,~}}           
\newcommand{\ie}{{\em i.e.,~}}           
\newcommand{\etc}{{\em etc.~}}         

\journal{A Journal}

\begin{document}

\begin{frontmatter}
\title{Joint Prediction and Time Estimation of COVID-19 Developing Severe Symptoms using  Chest CT Scan}

\author[label1]{Xiaofeng Zhu\corref{label5}}
\author[label2]{Bin Song\corref{label5}}
\author[label3]{Feng Shi}
\author[label3]{Yanbo Chen}
\author[label1]{Rongyao Hu}
\author[label1]{Jiangzhang Gan}
\author[label3]{Wenhai Zhang}
\author[label3]{Man Li}
\author[label3]{Liye Wang}
\author[label3]{Yaozong Gao}
\author[label4]{Fei Shan\corref{label6}}
\author[label3]{Dinggang Shen\corref{label6}}
\address[label1]{Center for Future Media  and school of computer science and technology, University of Electronic Science and Technology of China, Chengdu 611731, China.}
\address[label2]{Department of Radiology, Sichuan University West China Hospital, Chengdu 610041, China.}
\address[label3]{Department of Research and Development, Shanghai United Imaging Intelligence Co., Ltd., Shanghai 200232, China.}
\address[label4]{Department of Radiology, Shanghai Public Health Clinical Center, Fudan University, Shanghai, China.}
\cortext[label5]{X. Zhu (seanzhuxf@gmail.com) and B. Song (anicesong@vip.sina.com) contributed equally to this work.}
\cortext[label6]{Corresponding authors: F. Shan (shanfei\_2901@163.com) and D. Shen (Dinggang.Shen@gmail.com).}

\begin{abstract}
With the rapidly worldwide spread  of  Coronavirus  disease (COVID-19), it is of great importance to  conduct early diagnosis of COVID-19 and predict the time that patients might convert to the severe stage, for designing effective treatment plan and reducing  the clinicians' workloads.
In this study, we propose a joint classification and regression method to determine whether the patient would develop severe symptoms in the later time, and if yes, predict the possible conversion time that the patient would spend to convert to the severe stage.
To do this, the proposed method takes into account  1) the weight for each  sample to reduce the outliers' influence and explore the problem of imbalance classification, and 2)  the  weight for each feature via a sparsity regularization term to remove the redundant features of high-dimensional data and learn the shared information across the classification task and the regression task.
To our knowledge, this study is the first work to predict the disease progression and the conversion time, which could help clinicians to deal with the potential severe cases in time or even save the patients' lives.
Experimental analysis was conducted on a real data set from two hospitals with 422 chest computed tomography (CT) scans, where 52 cases were converted to severe on average 5.64 days and 34 cases were severe at admission. Results show  that  our  method  achieves the best classification (\eg 85.91\% of accuracy) and regression (\eg 0.462 of the correlation coefficient) performance, compared to all comparison methods. Moreover, our proposed method yields 76.97\% of accuracy for predicting the severe cases, 0.524 of the correlation coefficient, and 0.55 days difference for the converted time.
\end{abstract}


\end{frontmatter}

\section{Introduction}
Coronavirus disease (COVID-19) is an infectious disease caused by the severe-acute respiratory symptom Coronavirus 2 (SARS-Cov2), and has resulted in more than 1.3 million confirmed cases and about 75k deaths worldwide as of April 7, 2020 \cite{JHU}.
Due to its rapid spread, infection cases are increasing with high fatality rate (\eg up to 5.7\%).  Statistical models and machine learning methods have been developing to analyze the transmission dynamics and conduct early diagnosis of COVID-19 \cite{del2020covid,li2020early}.
For example,  Wu \etal employed the susceptible-exposed-infectious-recovered  model to estimate the number of the COVID-19 cases in Wuhan based on the number of exported COVID-19 cases which moved from Wuhan to other cities in China  \cite{wu2020nowcasting}.

The  increasing number of  confirmed COVID-19 cases  results in the lack of the clinicians and the increase of the clinicians' workloads. Many laboratory techniques have been used to confirm  the suspected COVID-19 cases by clinicians \cite{jung2020real}, including real-time reverse transcription polymerase chain reaction (RT-PCR) \cite{corman2020diagnostic,ai2020correlation}, non-PCR tests (\eg isothermal nucleic acid amplification technology \cite{craw2012isothermal}), non-contrast chest computed tomography (CT) and radiographs \cite{lee2020covid},  and so on.
It is well known that manually detection is time-consuming and increases the  infection risk of the clinicians \cite{kong2020chest}. Moreover, laboratory tests  are usually prohibited for all suspected cases due to the limitation of the test kits \cite{kong2020chest,ng2020imaging}. Also, RT-PCR has been widely used to confirm COVID-19, but  easily results in low sensitivity \cite{chaganti2020quantification}.
As a good alternative, artificial intelligence techniques on available data from laboratory tests have been playing important roles on the confirmation and follow-up of COVID-19 cases.
For example, Alom \etal employed  the inception  residual  recurrent convolutional  neural  network (CNN) and  transfer  learning  on X-ray and CT scan  images to detect COVID-19 and   segment  the infected regions of COVID-19 \cite{alom2020covidmtnet}.
Ozkaya \etal first applied CNN to  fuse and rank deep features for the early detection of COVID-19, and then used support vector machine (SVM) to conduct binary classification using the obtained deep features \cite{ozkaya2020coronavirus}.

While imaging data is playing an important role in the diagnosis  of  all kinds of pneumonia diseases including COVID-19 \cite{shi2020large},  CT has been applied to help monitor imaging changes and measure the disease severity \cite{zhao2020relation,chaganti2020quantification}.
For example, Chaganti \etal designed an automated system to  quantify the abnormal tomographic patterns appeared in COVID-19 \cite{chaganti2020quantification}.
Li \etal employed  CNNs with the imaging features of radiographs and CT images for identifying  COVID-19  \cite{li2020artificial}.

In this work, we investigate a new  early diagnosis method to predict whether the mild confirmed cases (\ie non-severe cases) of COVID-19 would develop severe symptoms in later time and estimate the time interval.
However, it is challenging due to many issues, such as small infected lesions in the chest CT scan at the early stage, appearances similar to other pneumonia,  the data set with high-dimensional features and small-sized samples, and imbalanced group distribution.

First, the infected lesions in the chest CT scan at the early stage are usually small and their appearances are quite similar with that of other pneumonia.
Given the early stage of COVID-19 with minor imaging signs, it is difficult to predict its future progression status. Conventional severity assessment methods  can easily distinguish a severe sign of the image from the mild  sign, since the changes of CT data are correlated with disease severity, \eg the lung involvement and abnormalities increase while the symptoms become severe. However, the infected volume of the non-severe COVID-19 cases is usually mild. For example, Guan \etal showed that  84.4\% of non-severe patients had mild symptoms  and more than 95\% severe cases had severe symptoms on CT changes \cite{guan2020clinical}. On the other hand, the clinicians have few prior knowledge about whether or when the non-severe cases convert to severe cases, so  early diagnosis and conversion time prediction could reduce the clinicians' workloads or even save patients' levies.

Second, the collected data set usually has a small number of samples (\ie small-sized samples) and high-dimensional features. Due to all kinds of reasons, such as data protection, data security, and the scenario of acute infectious diseases, a small number of subjects are available for early diagnosis of COVID-19. The limited  samples are difficult to build an effective artificial intelligence model. Moreover, high-dimensional features for each imaging data are often extracted, by considering to capture the comprehensive changes of the disease. Hence, both of scenarios often result in the issue of over-fitting and the issue of curse of dimensionality \cite{hu2019robust,zhu2014novel}.

Third, the class or group distribution of the data set is generally imbalanced.  In particular, the number of severe cases is much smaller than the number of non-severe cases, \eg 20\% reported\footnote{https://www.webmd.com/lung/news/20200324/the-other-side-of-covid-19-milder-cases-recovery\#1.\\https://www.businessinsider.com/coronavirus-80-percent-cases-are-mild-2020-2.}. Such a scenario poses a challenge for  most of classification methods because they were designed under the assumption of an equal number of samples for each class \cite{adeli2019logistic}. As a result, previous classification  techniques output poor predictive performance, especially for the minority class \cite{adeli2019logistic,zhu2019PR}.

In this work, we propose a novel joint regression and classification method to identify the severe COVID-19 cases from the non-severe cases and predict the conversion time from a non-severe case to the severe case in a unified framework. Specifically, we employ the logistic regression for the binary classification task and the linear regression for the regression task. Moreover, we  employ an $\ell_{2,1}$-norm regularization term on both the classification coefficient matrix and the regression coefficient matrix to consider the correlation between the two tasks as well as select the useful features for disease diagnosis and conversion time prediction. We further design a novel method to learn the weights of the samples, \ie automatically learning the weight of each sample so that  the important samples have large weights and the unimportant samples have small or even  zero weights. Moreover, the samples with zero weights in both the majority class and the minority class are excluded to the process of the construction of  the joint classification and regression, thus the problem of imbalance classification can be solved.

Different from previous literature, the contribution of our proposed method is listed as follows.

First, our method considers imbalance classification, feature selection, and sample weight in the same framework. Moreover, our method takes into account the correlation between the classification task and the regression task. In the literature,  few study has focused on exploring the above issues simultaneously. For example, the studies separately conduct feature selection  \cite{zhu2014novel} and sample weight \cite{hu2019robust,zhu2019eff} and Zhu \etal conduct joint classification and regression \cite{zhu2014novel}. Recently, a few studies simultaneously conduct feature selection and sample selection \cite{adeli2019logistic,adeli2018semi,hu2019robust}.

Second, in the literature, a few machine learning methods have proposed to conduct the diagnosis of COVID-19 disease. For example, Tang \etal employed random forest to detect the severe cases from the confirmed cases based on the CT scan data  \cite{tang2020severity}. Shi \etal conducted the same task by a two-step strategy, \ie automatically categorizing all subjects  into groups followed by random forests in each group for classification \cite{shi2020large}. However, previous literature did not take into account any of the above issues. Recently, deep learning techniques \cite{alom2020covidmtnet,ozkaya2020coronavirus,li2020artificial} have been employed to conduct early diagnosis of COVID-19, lacking the interpretability. To our knowledge, this is the first study simultaneously detecting the severe cases and predicting the conversion time, which have widely applications because the severe cases could endanger patients' lives and correctly predicting of the conversion time makes the clinicians take care the patients early or even save the patients' lives.

\renewcommand{\multirowsetup}{\centering}
\begin{table}[!t]
\centering
\caption{Demographic information of all subjects. The numbers in parentheses denote the number of subjects in each class. It is noteworthy that 52 cases were converted to severe on average 5.64 days and 34 cases were severe at admission.}
\vspace{-2mm}
\footnotesize{
\begin{tabular}{|c|c|c|}
\hline
\multirow{2}{*}{} &  Severe cases        & Non-severe cases    \\
                  &  (86)                & (322)  \\ \hline
Female/male       &  35/51               &   160/162     \\\hline
Age               &  55.43 $\pm$ 16.35   &    49.30 $\pm$ 15.70 \\    \hline
\end{tabular}
}
\label{tab0}
\vspace{-5mm}
\end{table}

\section{Materials and image preprocessing}
This study investigated the chest CT images of 422 confirmed COVID-19 patients. The demographic information is summarized in Table \ref{tab0}.
If a patient has multiple scans over time, the first scan is used.
All CT images  was provided by Shanghai Public Health Clinical Center and Sicuan University West China Hospital.
Informed consents were waived and all  private information of patients was anonymized. Moreover, the ethics of committees of these two institutes approved the protocol of this study.

All patients were confirmed by the national centers for disease control (CDC) based on the positive new Coronavirus nucleic acid antibody. Moreover, patients with large motion artifacts or pre-existing lung cancer conditions on the CT scans were excluded from this study.

\subsection{Image acquisition parameters}
All patients underwent thin-section CT scan by the scanners including SCENARIA 64 from Hitachi, Brilliance 64 from Philips, uCT 528 from United Imaging. The CT protocol is listed as follows: kV: 120, slice thickness: 1-1.5 mm, and breath hold at full inspiration. More details about both the image acquisition and the image pre-processing  can be found in  \cite{shan2020lung,shi2020large}. Moreover, we used the mediastinal window (with window width 350 hounsfield unit (HU) and window level 40 HU) and the lung window (with window width 1200 HU and window level-600 HU) for reading analysis.

\begin{figure}[!t]
 \begin{center}
  \vspace{-4mm}
  \subfigure {\scalebox{0.6} {\includegraphics{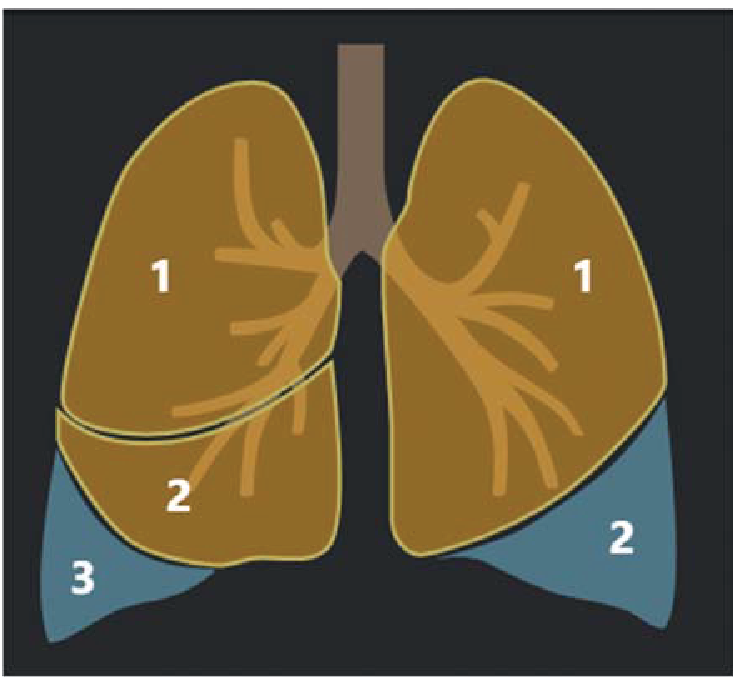}}}
  \subfigure {\scalebox{0.6} {\includegraphics{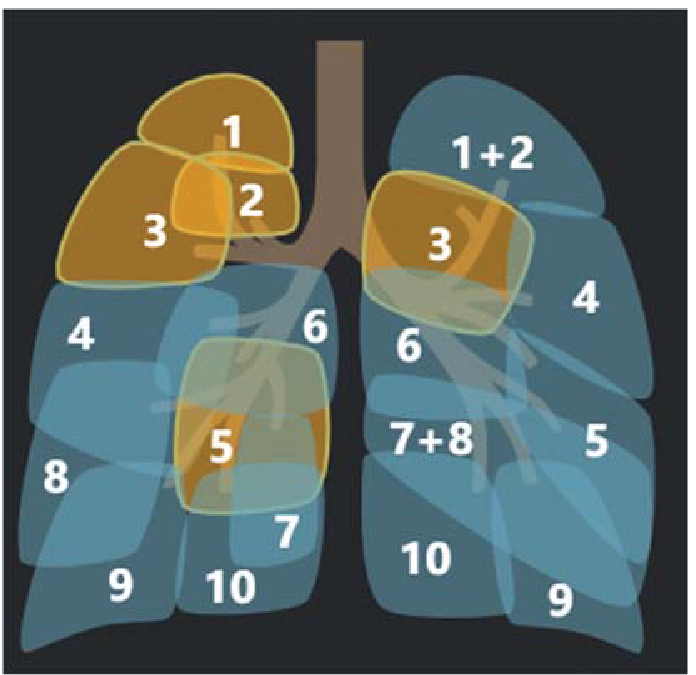}}}
  \caption{The visualization of the segments of a chest CT scan image, \ie 5 lung lobes (left) and 18 pulmonary segments.}
  \label{fig0}
 \end{center}
 \vspace{-7mm}
\end{figure}

\subsection{Image pre-processing}
We  utilized the disease characteristics, \ie infection locations and spreading patterns, to extract handcrafted features of each COVID-19 chest CT image. To do this, we used the  COVID-19  chest  CT  analysis  tool  developed  by Shanghai  United Imaging  Intelligence  Co.  Ltd.,  and followed the literature \cite{shan2020lung} to calculate the  quantitative features.

First,  the  COVID-19  chest  CT  analysis  tool designed a deep learning method named  VB-net to automatically segment infected lung regions and lung fields bilaterally.
The infected lung regions were mainly related to manifestations of pneumonia, such as mosaic sign, ground glass opacification (GGO), lesion-related signs, and interlobular septal thickening.

Second, after the segmentation process, the  lung fields include the left lung and the right lung, 5 lung lobes, and 18 pulmonary segments, as shown in Figure \ref{fig0}. Specifically, the left lung included superior lobe and inferior lobe, while the  right  lung included superior lobe, middle lobe, and inferior lobe. Moreover,  the left lung has 8 pulmonary segments and the right lung has 10 pulmonary segments. As a result, we had 26 regions of interest (ROIs) for each CT images.

Third, we partitioned each segment to five parts based on the HU ranges, \ie $HU_{[-\infty, -700]}$, $HU_{[-700, -500]}$, $HU_{[-500, -200]}$, $HU_{[-200, 50]}$, and $HU_{[50, \infty]}$. Specifically,
$HU_{[-\infty, -700]}$ indicates the parts with the HU range between $-\infty$ and -700,
$HU_{[-700, -500]}$ indicates the parts with the HU range between -700 and -500,
$HU_{[-500, -200]}$ indicates the parts with the HU range between -500 and -200,
$HU_{[-200, 50]}$ indicates the parts with the HU range between -200 and 50, and
$HU_{[50, \infty]}$ indicates the parts with the HU range between 50 and $\infty$.
As a result, each CT image was partitioned to 130 parts (\ie $390 = 26 * 5$).

In this study, we extracted three kinds of handcrafted features from each part, \ie density feature, volume feature, and mass feature.
Specifically, we  obtained the volume feature as the total volume of infected region and the density feature by calculating the averaged HU value within the infected region. We further followed \cite{song2014volume} to define the mass feature to simultaneously reflect the volume and density of subsolid nodule  because the mass feature has demonstrated to have  potentially superior reproducibility to 3D volumetry, \ie  $Mass =(Density + 1000) \times Volume \times 0.001$.

Finally, each CT image is represented by 390D handcrafted features in this study.

\section{Method}
In this paper, we denote matrices, vectors, and scalars, respectively, as boldface uppercase letters, boldface lowercase letters, and normal italic letters.
Specifically, we denote a matrix as $\mathbf{X} = [\mathbf{x}_{1}, ...,\mathbf{x}_{n}]$. The \emph{i}-th row and \emph{j}-th column of $\mathbf{X}$ are denoted as $\mathbf{x}^i$ and $\mathbf{x}_j$, respectively. We further denote the Frobenius norm and the $\ell_{2,1}$-norm  of a matrix $\mathbf{X}$ as $\|\mathbf{X}\|_F = \sqrt{\sum_i\|\mathbf{x}^i\|_2^2} = \sqrt{\sum_j\|\mathbf{x}_j\|_2^2}$ and
$\|\mathbf{X}\|_{2,1} = \sum_i\|\mathbf{x}^i\|_2 = \sum_i{\sqrt{\sum_j x_{ij}^2}}$, respectively. We also denote the transpose operator, the trace operator, and the inverse of a matrix $\mathbf{X}$ as $\mathbf{X}^T$, $tr(\mathbf{X})$, and $\mathbf{X}^{-1}$, respectively.

\subsection{Sparse logistic regression}
In the classification problem, given the feature matrix $\mathbf{X} \in \mathbb{R}^{d \times n}$ including \emph{n} samples $\mathbf{x}_i \in \mathbb{R}^{d}$ ($i = 1, ..., n$) and their corresponding labels  $y_i \in \{-1, +1\}$, the logistic regression is employed to distinguish the severe cases (\ie $y_i = +1$) from the confirmed COVID-19 cases (\ie $y_i = -1$).
Specifically, by denoting $\mathbf{w} \in \mathbb{R}^{d}$ as the coefficient vector,
the logistic loss function is defined as
\begin{equation}
\min \limits_{\mathbf{w}} ~ \sum \limits_{i = 1}^{n}   log(1+exp(-y_i(\mathbf{w}^T\mathbf{x}_i))) + \lambda \|\mathbf{w}\|_2^2
\label{eq1}
\end{equation}
where $\lambda_1$ is a tuning parameter, and the $\ell_2$-norm regularization term on the coefficient vector $\mathbf{w}$ is used to control the complexity of the logistic regression. Eq. (\ref{eq1}) conducts the classification task without taking into account the issues, such as feature selection, sample weight, and imbalance classification.

First, in real applications, clinicians have prior knowledge on the regions  of the CT scan data which are possible related to the disease, but we cannot only extract the features from these regions because they may cooperate with other regions to  influence the disease. As a result, we extract the features from all imaging data to obtain high-dimensional data, which captures the comprehensive changes of confirmed COVID-19 cases but increases the store  and computation costs as well as easily results in the issue of curse of dimensionality \cite{zhu2019PR}.
To address this issue, we design machine learning models to automatically recognize the features related to the  disease by taking into account the correlation of the features.
Specifically,  we replace the $\ell_2$-norm on the coefficient vector $\mathbf{w}$ (\ie $\|\mathbf{w}\|_2^2$) by the $\ell_1$-norm  on on the coefficient vector $\mathbf{w}$ (\ie $\|\mathbf{w}\|_1$ and $\|\mathbf{w}\|_1  = \sum_{i = 1}^{d} |w_i|$), which outputs sparse elements to make the corresponding features (\ie the rows in $\mathbf{X}$) not involving the classification task, \ie
\begin{equation}
\min \limits_{\mathbf{w}} ~ \sum\limits_{i=1}^{n}  log(1+exp(-y_i(\mathbf{w}^T\mathbf{x}_i))) + \lambda_2 \|\mathbf{w}\|_1
\label{eq2}
\end{equation}
where $\lambda_2$ is a tuning parameter.

\subsection{Balanced and sparse logistic regression}
In binary classification, the issue of imbalance classification easily results in the classification results bias to the majority class, \ie outputting high false negatives.
In the literature, both re-sampling methods and cost-sensitive learning methods \cite{zhu2019eff} have been used for solve the issue of imbalance classification.

Recently, robust loss functions has been widely designed to reduce the influences of outliers by taking into account the sample weight in robust statistics \cite{zhu2019PR}.
Specifically, robust loss functions use a weight vector $\bm{\alpha} \in \mathbb{R}^{1 \times n}$ to automatically output small weights to the samples with large estimation errors and large weights to the samples with small estimation error. As consequence, the samples with large estimation errors are regarded as outliers and their influences are reduced.
In the literature, a number of robust loss functions have been developed, including  $\ell_{1}-\ell_{2}$ function, Cauchy function, and Geman–McClure estimator, \etc \cite{hu2019robust,zhu2019HQD}.
However, the robust loss function was not designed to explore the issue of  imbalance classification.

In this paper,  motivated by the self-paced learning assigning weights to the samples, we propose a new method to assign a weight to each sample as well as to solve the problem of imbalance classification.
By regarding that different samples have different contributions to the construction of the classification model, our method expect to assign large weights to the important samples  and small weights to the unimportant samples. Moreover, by regarding of the problem of imbalance classification, our method expect to set different numbers of zero weights  to different classes so that there is  a balance of the sample number between the positive class and the negative class. To do this, we employ an $\ell_0$-norm constraint on the weight vector $\bm{\alpha}$ to have the following loss function:
\begin{eqnarray}
\begin{array}{l}
\min \limits_{\bm{\alpha}, \mathbf{w}}~  \sum\limits_{i=1}^{n} {\alpha}_i log(1+exp(-y_i(\mathbf{w}^T\mathbf{x}_i))) + \lambda_2 \|\mathbf{w}\|_1 \\
~~~~ s.t., ~ \|\bm{\alpha}_{-1}\|_0 = k_-  ~and~ \|\bm{\alpha}_{+1}\|_0 = k_+
\end{array}
\label{eq3}
\end{eqnarray}
where $\bm{\alpha}_{-1}$ indicates the weight set of all negative samples and $\bm{\alpha}_{+1}$ indicates the weight set of all positive samples.  The constraint  `$\|\bm{\alpha}_{-1}\|_0 =  k_-$' indicates that the number of non-zero elements in the negative class is $k_-$. Specifically, after receiving the estimation value for each sample, \ie $log(1 + exp(-(y_i\mathbf{w}^T\mathbf{x}_i)))$, we first sort the estimation values of all samples in the same class with an increase order. We then keep the original weights to the weights of the $k_-$ negative samples with the smallest estimation values and 0 to the weights of the left negative samples. In this way, either the negative  samples or the positive samples with zero weights will not involve the process of the classification model.

Our method in Eq. (\ref{eq3}) has at least two advantages, \ie automatically selecting important samples (\ie reducing the influence of the outliers) to learn the classification model and adjusting the number of selected samples for each class by tuning the values of $k_-$ and $k_+$ (\eg $k_- = k_+$) to solve the imbalance classification problem.
In particular, our method in Eq. (\ref{eq3}) employs the $\ell_0$-norm constraint for each class to output exactly predefined non-zero elements. On the contrary, self-paced learning uses the $\ell_1$-norm constraint for all samples or other robust loss functions  \cite{zhu2019PR,hu2019robust} to  estimate the sample weight without guaranteeing the exact number of non-zero elements. As a result, compared to self-paced learning only considering the sample weight to reduce the influence of outliers, our method takes into account  the sample weight to remove outliers not to involve the process of the  model construction as well as the problem of imbalance classification.

\subsection{Joint logistic regression and linear regression}
Besides distinguishing the severe cases from the non-severe cases, predicting the time converting a non-severe case to  a severe case is also important because it may be related to the patients' lives. To do this, a naive solution is to separately conduct a classification task to diagnose the severe cases and  a regression task to predict the conversion time. Obviously, the separate strategy ignores the correlation among two tasks.
In this paper, by regarding the prediction of conversion time as a regression task, we define a ridge regression to linearly characterize the correlation between the feature matrix $\mathbf{X}$ and the vector of conversion time  $\mathbf{z} \in \mathbb{R}^{1 \times n}$ by
\begin{eqnarray}
\min \limits_{\mathbf{v}} ~ \|\mathbf{v}^T\mathbf{X} - \mathbf{z}\|_2^2 + \lambda_3\|\mathbf{v}\|_2^2
\label{eq4}
\end{eqnarray}
where $\mathbf{v} \in \mathbb{R}^d$  is the coefficient vector for the regression task  and $\lambda_3$ is a tuning parameter.

Similar to the classification task in Eq. (\ref{eq3}), the regression task in  Eq. (\ref{eq4}) still needs to consider the issues, such as feature selection, sample weight, and imbalance classification. Moreover, in this study, we conduct joint classification and regression (\ie multi-task learning) by simultaneously considering a classification task and a regression task in the same framework. We expect that each task could obtain information from another task so that the model effectiveness of each of them can be improved by the  shared information.
Specifically,  we employ the $\ell_{2,1}$-norm regularization term with respect to both the variable $\mathbf{w}$ and the variable $\mathbf{v}$ to obtain the following objective function:
\begin{eqnarray}
\begin{array}{l}
\min \limits_{\bm{\alpha},  \mathbf{w}, \mathbf{v}, \bm{\beta}} ~ \sum\limits_{i=1}^{n} {\alpha}_i log(1+exp(-y_i(\mathbf{w}^T\mathbf{x}_i)))\\
~~~~ ~~~~~~ + \gamma\|\bm{\beta}(\mathbf{v}^T\mathbf{X}  - \mathbf{z})\|_2^2
 + \lambda \|[\mathbf{w},\mathbf{v}]\|_{2,1}\\
~~~~ s.t., ~ \|\bm{\alpha}_{-1}\|_0 = k_-  ~and~ \|\bm{\alpha}_{+1}\|_0 = k_+,\\
~~~~ ~~~~~~  \|\bm{\beta}_{-1}\|_0 = k_-  ~and~ \|\bm{\beta}_{+1}\|_0 = k_+
\end{array}
\label{fobj1}
\end{eqnarray}
where $\bm{\beta} \in \mathbb{R}^{1 \times n}$ is the sample weight vector for the regression task and $[\mathbf{w}, \mathbf{v}] \in \mathbb{R}^{d \times 2}$ . $\gamma$ and $\lambda$ are tuning parameters.
$\|[\mathbf{w},\mathbf{v}]\|_{2,1}$ indicates that the selected features are obtained by the classification and regression model. Moreover, the selected features are their  shared or common information benefiting each of them \cite{evgeniou2004regularized}.

Eq. (\ref{fobj1}) needs a  tuning parameter $\gamma$ to have a magnitude or importance trade-off for two tasks. However, the process of tuning parameter is time-consuming and needs prior knowledge. In this work, we use a squared root operator on the second term of Eq. (\ref{fobj1}) to automatically obtain their weights. It is noteworthy that we keep the parameter $\lambda$ to be tuned because it controls the sparsity of the term $\|[\mathbf{w},\mathbf{v}]\|_{2,1}$ and the sparsity will be changed based on the data distribution \cite{evgeniou2004regularized,zhu2017graph}. Hence, the final objective function of our proposed joint classification and regression method is:
\begin{eqnarray}
\begin{array}{l}
\min \limits_{\bm{\alpha},  \mathbf{w}, \mathbf{v}, \bm{\beta}} ~ \sum\limits_{i=1}^{n} {\alpha}_i log(1+exp(-y_i(\mathbf{w}^T\mathbf{x}_i)))\\
~~~~ ~~~~~~ + \sqrt{\sum \limits_{i = 1}^{n} \beta_i \|\mathbf{v}^T\mathbf{x}_i  - z_i\|_2^2}
 + \lambda \|[\mathbf{w},\mathbf{v}]\|_{2,1}\\
~~~~ s.t., ~ \|\bm{\alpha}_{-1}\|_0 = k_-  ~and~ \|\bm{\alpha}_{+1}\|_0 = k_+,\\
~~~~ ~~~~~~  \|\bm{\beta}_{-1}\|_0 = k_-  ~and~ \|\bm{\beta}_{+1}\|_0 = k_+
\end{array}
\label{fobj}
\end{eqnarray}

To solve the optimization problem in Eq. (\ref{fobj}), \ie optimizing the variables $\mathbf{v}$ and $\bm{\beta}$,   we compute the derivatives of the square root in Eq. (\ref{fobj}) and obtain the following formulation
\begin{subequations}
\begin{numcases}{}
\min \limits_{\bm{\alpha},  \mathbf{w},  \mathbf{v}, \bm{\beta}} ~ \sum\limits_{i=1}^{n} {\alpha}_i log(1+exp(-y_i(\mathbf{w}^T\mathbf{x}_i))) \label{fobj11}\\
 ~~~~~~ + \gamma \sum \limits_{i = 1}^{n} \beta_i \|\mathbf{v}^T\mathbf{x}_i - z_i\|_2^2 + \lambda \|[\mathbf{w},\mathbf{v}]\|_{2,1} \nonumber\\
~~~~ s.t., ~ \|\bm{\alpha}_{-1}\|_0 = k_-  ~and~ \|\bm{\alpha}_{+1}\|_0 = k_+,\nonumber\\
~~~~ ~~~~~~  \|\bm{\beta}_{-1}\|_0 = k_-  ~and~ \|\bm{\beta}_{+1}\|_0 = k_+\nonumber\\
 \gamma = \frac{1}{2\sqrt{\|\bm{\beta}(\mathbf{v}^T\mathbf{X}  - \mathbf{z})\|_2^2}}. \label{fobj12}
\end{numcases}
\end{subequations}

The values of $\gamma$ in Eq. (\ref{fobj12}) is automatically obtained without the tuning process and can be regarded as the weight of the tasks. Specifically, if the prediction error is small, the value of $\gamma$ is large, \ie the regression task is more important than the classification task. Hence, the optimization of the value of $\gamma$ automatically balances the contributions of two tasks. As a result, the optimization of Eq. (\ref{fobj}) is changed to optimize Eq. (\ref{fobj12}).

\begin{algorithm}[tb]
\caption{The pseudo of our optimization method.}
\label{alg1}
\textbf{Input}: $\mathbf{X}\in \mathbb{R}^{d \times n}$, $\mathbf{y} \in \mathbb{R}^{1 \times n}$, $\mathbf{z}\in \mathbb{R}^{1 \times n}$, and $\lambda$.\\
\textbf{Output}: $\mathbf{w}\in \mathbb{R}^{d}$, $\mathbf{v}\in \mathbb{R}^{d}$, $\bm{\alpha} \in \mathbb{R}^{1 \times n}$, and $\bm{\beta} \in \mathbb{R}^{1 \times n}$.
\begin{algorithmic}[1]
\STATE Random initialize $\mathbf{D} \in \mathbb{R}^{d \times d}$;
\WHILE{$Eq.(\ref{fobj})$ \emph{not converge}}
\STATE Update $\mathbf{w}$ via Eq.(\ref{op4});
\STATE Update $\mathbf{v}$ via Eq.(\ref{op8});
\STATE Update $\mathbf{\alpha}$ via Eq.(\ref{op9});
\STATE Update $\mathbf{\beta}$ via Eq.(\ref{op10});
\STATE Update $\mathbf{D}$ by $[\mathbf{w},~\mathbf{v}]$;
\STATE Update $\gamma$ by Eq.(\ref{fobj12});
\ENDWHILE
\end{algorithmic}
\end{algorithm}
\vspace{-3mm}

\subsection{Optimization}
In this paper, we employ the alternating optimization strategy \cite{bezdek2003convergence}  to optimize the variables
$\mathbf{w}$, $\bm{\alpha}$,  $\mathbf{v}$, and $\bm{\beta}$, in Eq. (\ref{fobj11}). We list the pseudo of our optimization method in Algorithm \ref{alg1} and report the details as follows.

\textbf{(i) Update $\mathbf{w}$  by fixing $\bm{\alpha}$, $\bm{\beta}$ and $\mathbf{v}$}

After  other variables are fixed, the objective function with respect to $\mathbf{w}$ in Eq. (\ref{fobj11}) becomes
\begin{eqnarray}
\begin{array}{l}
\mathop{\min} \limits_{\mathbf{w}} \sum\limits_{i=1}^{n} {\alpha}_i log(1+exp(-y_i(\mathbf{w}^T\mathbf{x}_i)))
 + \lambda tr(\mathbf{w}^T\mathbf{D}\mathbf{w})
\end{array}
\label{op1}
\end{eqnarray}
where $\mathbf{D}\in \mathbb{R}^{d \times d}$ is a diagonal matrix and the value of its $i$-th diagonal element is $\frac{1}{2\|[\mathbf{w}, \mathbf{v}]^i\|_2}$.
To solve the imbalance classification problem, we set $k_- = k_+ = k$ to obtain $2k$ samples for the training process. Hence, Eq. (\ref{op1}) becomes
\begin{eqnarray}
\begin{array}{l}
 \mathop{\min} \limits_{\mathbf{w}} \sum\limits_{i=1}^{2k} {\alpha}_i log(1+exp(-y_i(\mathbf{w}^T\mathbf{x}_i))) + \lambda tr(\mathbf{w}^T\mathbf{D}\mathbf{w})
\end{array}
\label{op2}
\end{eqnarray}

By setting $y_i \in \{-1,1\}$ and $h_w(\mathbf{x}_i) = \frac{1}{1+exp(-(\mathbf{w}^T\mathbf{x}_i))}$,  Eq. (\ref{op2}) becomes
\begin{eqnarray}
\begin{array}{l}
\mathop{\min} \limits_{\mathbf{w}} \sum\limits_{i=1}^{2k} {\alpha}_i(- \frac{1+y_i}{2}log(h_w(\mathbf{x}_i))- \frac{1-y_i}{2}log(1-h_w(\mathbf{x}_i)))
+ \lambda tr(\mathbf{w}^T\mathbf{D}\mathbf{w})
\end{array}
\label{op3}
\end{eqnarray}

In this paper, we employ the Newton's method \cite{liu1989limited} to minimize Eq. (\ref{op3}) by the following update rules
\begin{eqnarray}
\begin{array}{l}
\mathbf{w} = \mathbf{w} -\mathbf{a}\mathbf{B}^{-1}
\end{array}
\label{op4}
\end{eqnarray}
where $\mathbf{a} \in \mathbb{R}^{d}$ and  $\mathbf{B} \in \mathbb{R}^{d \times d}$ are defined as
\begin{eqnarray}
\begin{array}{l}
\left\{
\begin{aligned}
\mathbf{a} = \sum\limits_{i=1}^{2k} {\alpha}_i(h_{w}(\mathbf{x}_i)- \frac{1+y_i}{2}) \mathbf{x_i} + 2 \lambda\mathbf{Dw} \\
\mathbf{B} = \sum\limits_{i=1}^{2k} {\alpha}_i h_{w}(\mathbf{x}_i)(1 - h_{w}(\mathbf{x}_i)) \mathbf{x}_i \mathbf{x}_i^T + 2\lambda\mathbf{D}
\end{aligned}
\right.
\end{array}
\label{op5}
\end{eqnarray}

\textbf{(ii) Update $\mathbf{v}$  by fixing $\bm{\alpha}$, $\bm{\beta}$ and $\mathbf{w}$}

The objective function with respect to $\mathbf{v}$ in Eq. (\ref{fobj11}) is
\begin{eqnarray}
\begin{array}{l}
\mathop{\min} \limits_{\mathbf{v}} \sum\limits_{i=1}^{n} {\beta}_i \|\mathbf{v}^T x_i - z_i\|_F^2 + \frac{\lambda}{\gamma} tr(\mathbf{v}^T\mathbf{D}\mathbf{v})
\end{array}
\label{op6}
\end{eqnarray}

By letting  $\mathbf{G} = [\beta_1\mathbf{x}_1, ..., \beta_n\mathbf{x}_n] \in \mathbb{R}^{d \times n}$ and $\mathbf{m} = [\beta_1{z}_1, ..., \beta_n{z}_n] \in \mathbb{R}^{1 \times n}$, we have
\begin{eqnarray}
\begin{array}{l}
\mathop{\min} \limits_{\mathbf{v}}   \|\mathbf{v}^T \mathbf{G}-\mathbf{m}\|_2^2 + \frac{\lambda}{\gamma} tr(\mathbf{v}^T\mathbf{D}\mathbf{v})
\end{array}
\label{op7}
\end{eqnarray}

Eq. (\ref{op7}) has a closed-form solution, \ie
\begin{eqnarray}
\begin{array}{l}
\mathbf{v} = (\mathbf{G}\mathbf{G}^T + \frac{\lambda}{\gamma}\mathbf{D})^{-1} \mathbf{Gm}
\end{array}
\label{op8}
\end{eqnarray}

\textbf{(iii) Update $\bm{\alpha}$ and $\bm{\beta}$  by fixing $\mathbf{v}$ and $\mathbf{w}$}

By denoting the estimation value of the \emph{i}-th sample as $l_1^i = log(1+exp(-y_i\mathbf{w}^T\mathbf{x}_i))$ ($i = 1, ...,n$), we sort the values $l_1^i$ ($i = 1, ...,n$) with an increase order for each class to denote the weight set of $k$ negative samples with the smallest estimation values as $I_- = \{\hat{\alpha}_{[1]}^-, ..., \hat{\alpha}_{[k]}^-\}$ (where $\hat{\alpha}_{[i]}^- < \hat{\alpha}_{[j]}^-$ if $[i] < [j]$)  and  the weight set of $k$ positive samples with the smallest estimation values as $I_+ = \{\hat{\alpha}_{[1]}^+, ..., \hat{\alpha}_{[k]}^+\}$ (where $\hat{\alpha}_{[i]}^+ < \hat{\alpha}_{[j]}^+$ if $[i] < [j]$), and then we have
\begin{eqnarray}
\begin{array}{l}
\hat{\alpha}_{[i]}=\left\{
\begin{aligned}
\alpha_{j} ~~~ [i] \in I_- \cup I_+ ~and~ [i] \bowtie j& \\
0~~~~~~~~~others
\end{aligned}
\right.
\end{array}
\label{op9}
\end{eqnarray}
where $\{1, ..., n\}$ is the index of the original order before the sorting and $\{[1], ..., [n]\}$ is the index of the increase order after the sorting. $[i] \bowtie j$ indicates that the \emph{j}-th index in the original order is matched with the [\emph{i}]-th index in the increase order.

By denoting the estimation value of the \emph{i}-th sample as $l_2^i = \|\mathbf{v}^T\mathbf{x}_i - z_i\|_2^2$ ($i = 1, ...,n$), we sort the values $l_2^i$ ($i = 1, ...,n$) with an increase order for each class to denote the weight set of $k$ negative samples with the smallest estimation values as $Q_- = \{\hat{\beta}_{[1]}^-, ..., \hat{\beta}_{[k]}^-\}$ (where $\hat{\beta}_{[i]}^- < \hat{\beta}_{[j]}^-$ if $[i] < [j]$)  and  the weight set of $k$ positive samples with the smallest estimation values as $Q_+ = \{\hat{\beta}_{[1]}^+, ..., \hat{\alpha}_{[k]}^+\}$ (where $\hat{\beta}_{[i]}^+ < \hat{\beta}_{[j]}^+$ if $[i] < [j]$), and then we have
\begin{eqnarray}
\begin{array}{l}
\hat{\beta}_{[i]}=\left\{
\begin{aligned}
\beta_{j} ~~~ [i] \in Q_- \cup Q_+ ~and~ [i] \bowtie j& \\
0~~~~~~~~~others
\end{aligned}
\right.
\end{array}
\label{op10}
\end{eqnarray}

\subsection{Convergence analysis} \label{sec24}
Algorithm \ref{alg1} involves five variables (\ie $\mathbf{w}$, $\bm{\alpha}$,  $\mathbf{v}$,  $\bm{\beta}$, and $\gamma$).
By denoting $\mathbf{w}^{t}$, $\bm{\alpha}^{t}$,  $\mathbf{v}^{t}$,  $\bm{\beta}^{t}$, and $\gamma^{t}$, respectively, as the $t$-th iteration  results of $\mathbf{w}$, $\bm{\alpha}$,  $\mathbf{v}$,  $\bm{\beta}$, and $\gamma$, we denote the objective function value of the $t$-th iteration of Eq. (\ref{fobj}) as $J(\mathbf{w}^{t}, \bm{\alpha}^{t},  \mathbf{v}^{t},  \bm{\beta}^{t}, \gamma^{t})$.

By fixing $\bm{\alpha}^{t}$,  $\mathbf{v}^{t}$,  $\bm{\beta}^{t}$, and $\gamma^{t}$, we employ the Newton's method to optimize $\mathbf{w}$, so we have
\begin{eqnarray}
\begin{array}{l}
J(\mathbf{w}^{t+1}, \bm{\alpha}^{t},  \mathbf{v}^{t},  \bm{\beta}^{t},  \gamma^{t})
 \leq J(\mathbf{w}^{t}, \bm{\alpha}^{t},  \mathbf{v}^{t},  \bm{\beta}^{t},  \gamma^{t})
\end{array}
\label{cp01}
\end{eqnarray}

The optimizations of the variables (\ie $\bm{\alpha}$,  $\mathbf{v}$,  $\bm{\beta}$, and $\gamma$) have closed-form solutions, so we have
\begin{eqnarray}
\begin{array}{l}
J(\mathbf{w}^{t+1}, \bm{\alpha}^{t+1},  \mathbf{v}^{t},  \bm{\beta}^{t},  \gamma^{t})  \leq J(\mathbf{w}^{t+1}, \bm{\alpha}^{t},  \mathbf{v}^{t},  \bm{\beta}^{t},  \gamma^{t})
\end{array}
\label{cp02}
\end{eqnarray}

\begin{eqnarray}
\begin{array}{l}
J(\mathbf{w}^{t+1}, \bm{\alpha}^{t+1},  \mathbf{v}^{t+1},  \bm{\beta}^{t},  \gamma^{t}) \leq J(\mathbf{w}^{t+1}, \bm{\alpha}^{t+1},  \mathbf{v}^{t},  \bm{\beta}^{t},  \gamma^{t})
\end{array}
\label{cp03}
\end{eqnarray}
\begin{eqnarray}
\begin{array}{l}
J(\mathbf{w}^{t+1}, \bm{\alpha}^{t+1},  \mathbf{v}^{t+1},  \bm{\beta}^{t+1},  \gamma^{t}) \leq J(\mathbf{w}^{t+1}, \bm{\alpha}^{t+1},  \mathbf{v}^{t+1},  \bm{\beta}^{t},  \gamma^{t})
\end{array}
\label{cp04}
\end{eqnarray}

\begin{eqnarray}
\begin{array}{l}
J(\mathbf{w}^{t+1}, \bm{\alpha}^{t+1},  \mathbf{v}^{t+1},  \bm{\beta}^{t+1},  \gamma^{t+1}) \leq J(\mathbf{w}^{t+1}, \bm{\alpha}^{t+1},  \mathbf{v}^{t+1},  \bm{\beta}^{t+1},  \gamma^{t})
\end{array}
\label{cp05}
\end{eqnarray}

By integrating Eqs. (\ref{cp01})-(\ref{cp05}), we obtain
\begin{eqnarray}
\begin{array}{l}
J(\mathbf{w}^{t+1}, \bm{\alpha}^{t+1},  \mathbf{v}^{t+1},  \bm{\beta}^{t+1},  \gamma^{t+1}) \leq J(\mathbf{w}^{t}, \bm{\alpha}^{t},  \mathbf{v}^{t},  \bm{\beta}^{t},  \gamma^{t})
\end{array}
\label{cp06}
\end{eqnarray}

According to Eq. (\ref{cp06}), the objective function values in Eq. (\ref{fobj}) gradually decrease with the increase of iterations until Algorithm \ref{alg1} converges. Hence, the convergence proof of Algorithm \ref{alg1} to optimize  Eq. (\ref{fobj}) is completed.

\begin{table}[!t]
\centering
\caption{The summarization of all methods. Note that, FS: feature selection, SW: sample weight, IMB: imbalance classification, CLASS: classification, and REG: regression.}
\footnotesize{
\begin{tabular}{|c|c|c|c|c|c|} \hline
Methods&FS&SW&IMB&CLASS&REG\\ \hline
SVM \cite{chang2011libsvm}          &&        &       &$\surd$& \\
L1SVM \cite{chang2011libsvm}          &$\surd$&        &       &$\surd$& \\
Random forest \cite{liaw2002classification} && &$\surd$ &$\surd$& \\
SFS \cite{adeli2019logistic}          &$\surd$&$\surd$ & $\surd$  &$\surd$& \\ \hline
Ridge regression      && & &           &$\surd$ \\
L1SVR \cite{chang2011libsvm}     &$\surd$& & &           &$\surd$ \\
Lasso \cite{tibshirani1996regression} &$\surd$& & &                    &$\surd$ \\
Random forest \cite{liaw2002classification} &$\surd$& &$\surd$ &&$\surd$ \\\hline
MSFS \cite{zhu2014novel}              &$\surd$& &              &$\surd$&$\surd$ \\
Proposed     &$\surd$&$\surd$ &$\surd$ &$\surd$&$\surd$ \\ \hline
\end{tabular}
}
\label{tab1}
\end{table}

\section{Experiments}
We experimentally evaluated our method, compared to  state-of-the-art classification and regression methods, on a real COVID-19 data set with chest CT scan data, in terms of binary classification performance and regression performance.

\subsection{Experimental setting}
We selected SVM and ridge regression, respectively, as the baseline methods for the classification task and the regression task. Other comparison methods include $\ell_1$-SVM (L1SVM) \cite{chang2011libsvm}, random forest  \cite{liaw2002classification}, sample-feature selection (SFS) \cite{adeli2019logistic},  $\ell_1$-SVR \cite{chang2011libsvm} (L1SVR), least absolute shrinkage and selection operator (Lasso) \cite{tibshirani1996regression}, and matrix-similarity feature selection (MSFS) \cite{zhu2014novel}. We summarize the details of all comparison methods in Table \ref{tab1}. It is noteworthy that random forest can be used for feature selection,  sample selection, and imbalance classification. However, in this study, we only used random forest to consider the problem of imbalance classification.

In our experiments, we repeated the 5-fold cross-validation scheme 20 times for all methods to report the average results as the final results. In the model selection, we set $\lambda \in \{10^{-3}, 10^{-2},  ..., 10^{3}\}$ in Eq. (\ref{fobj}), and fixed $k = 50$ for solving the problem of imbalance classification for our method. We followed the literature \cite{chang2011libsvm,liaw2002classification,adeli2019logistic,tibshirani1996regression,zhu2014novel} to set the parameters of the comparison methods so that they outputted the best results.

The evaluation metrics include accuracy, specificity, sensitivity, and area under the ROC curve (AUC) for the classification task, as well as correlation coefficient (CC) and root mean square error (RMSE) for the regression task.

\begin{figure}[!t]
 \begin{center}
  \subfigure [Classification results]{\scalebox{0.275} {\includegraphics{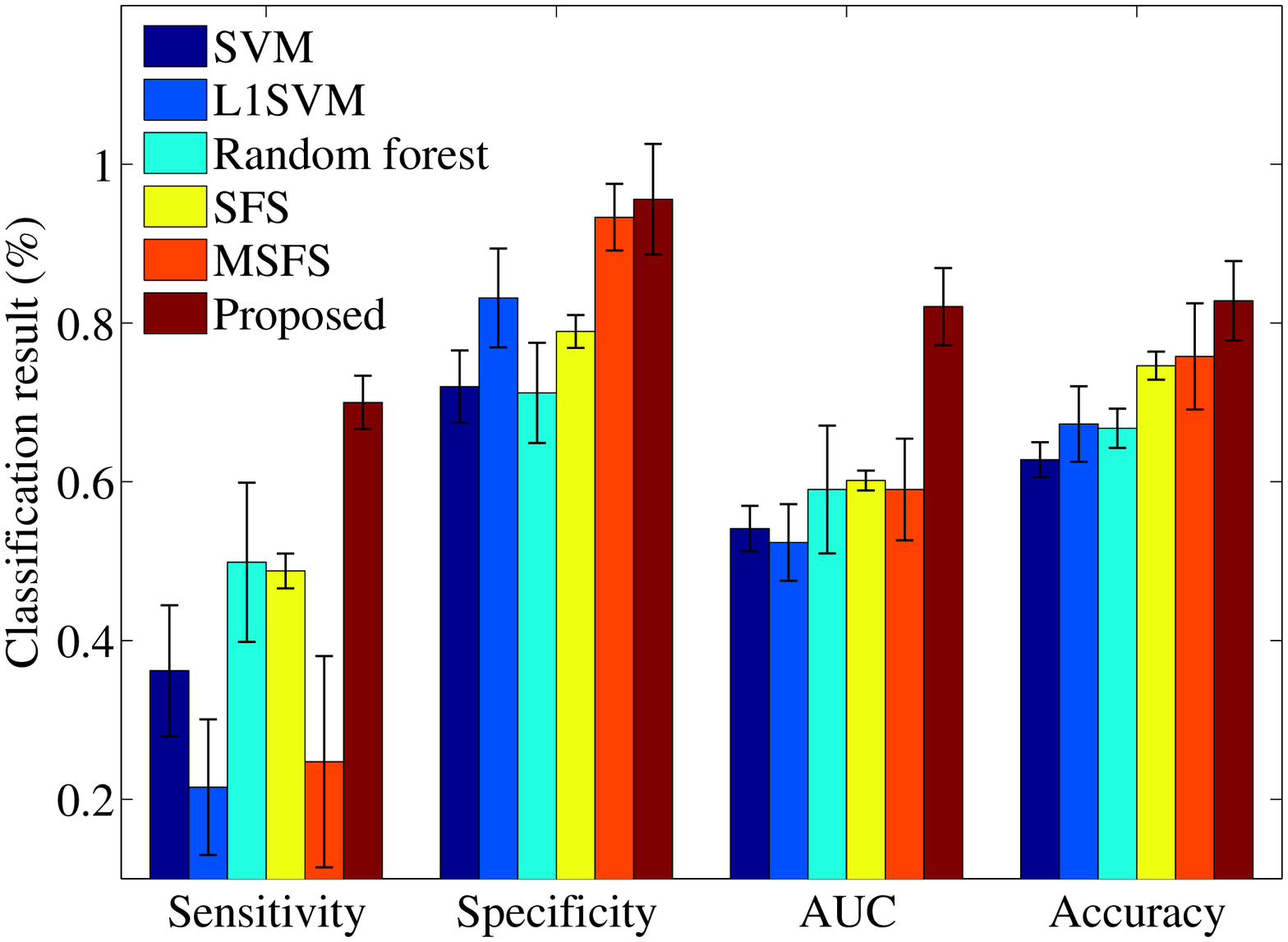}}}
  \subfigure [ROC curves]{\scalebox{0.425} {\includegraphics{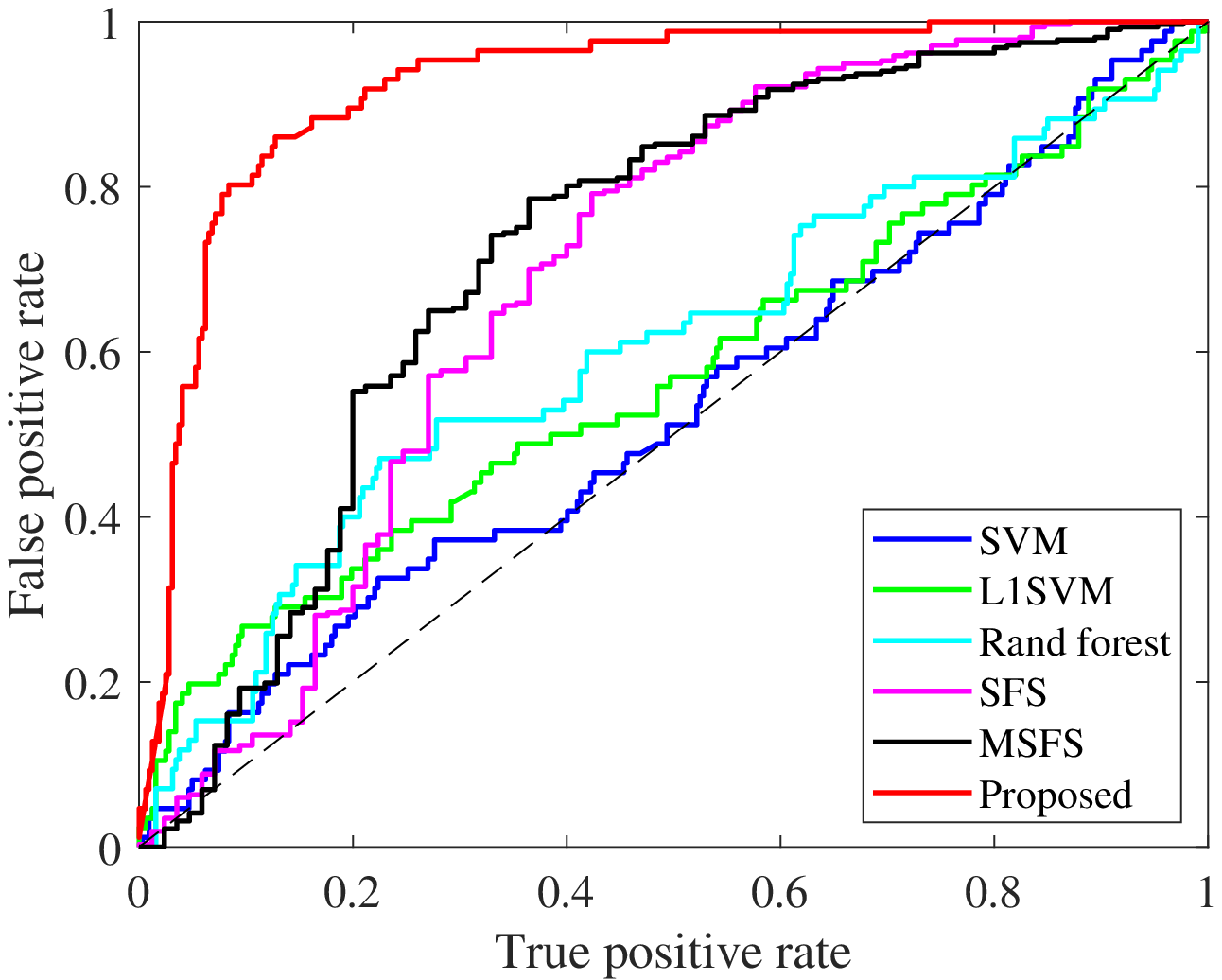}}}
  \caption{Classification results and ROC curves of all methods.}
  \label{fig1}
 \end{center}
\end{figure}

\eat{
\begin{figure}[!t]
 \begin{center}
  \subfigure {\scalebox{0.5} {\includegraphics{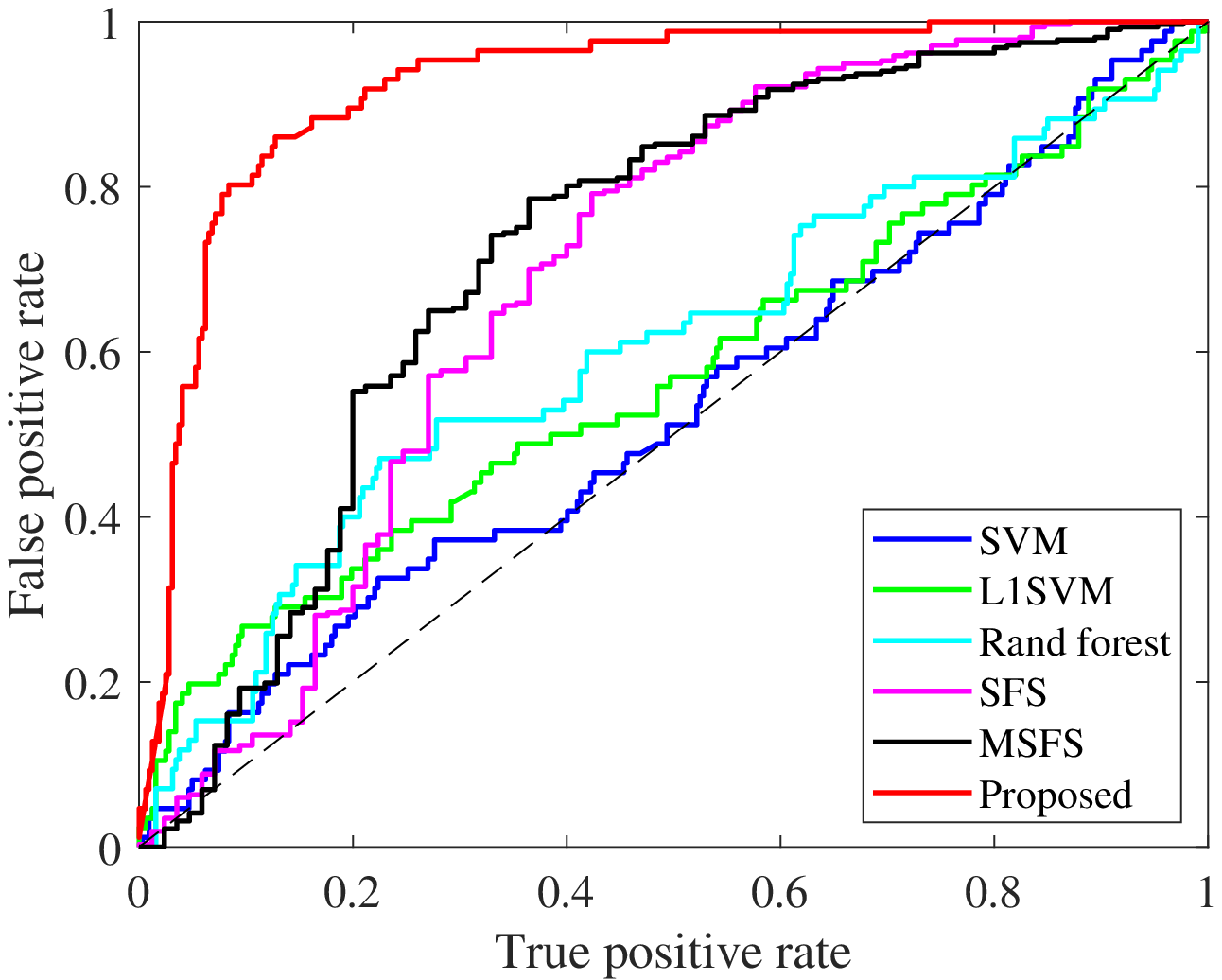}}}
  \vspace{-3mm}
  \caption{ROC curves of all methods.}
  \label{fig2}
 \end{center}
 \vspace{-5mm}
\end{figure}
}

\begin{table}[!tb]
	\centering
	\caption{Classification results (\%) of three methods. Proposed W/O Regression indicates Eq. (\ref{eq3}).}
	\centering
	{\scriptsize
		\begin{tabular}[c]{|c|c|c|c|}\hline
Methods     &SFS  \cite{adeli2019logistic}  & Proposed w/o Regression  &Proposed    \\\hline
Accuracy    &78.18 $\pm$ 3.71 &83.25 $\pm$ 2.44 &\textbf{85.69} $\pm$ 2.20 \\ \hline
Sensitivity &50.65 $\pm$ 6.33 &70.73 $\pm$ 3.36 &\textbf{76.97} $\pm$ 3.36 \\ \hline
Specificity &86.31 $\pm$ 2.69 &86.60 $\pm$ 3.45 &\textbf{88.02} $\pm$ 1.45 \\ \hline
AUC         &73.88 $\pm$ 6.66 &81.74 $\pm$ 3.30 &\textbf{85.91} $\pm$ 2.27 \\ \hline
\end{tabular}
	}
\label{tab2}
\end{table}

\subsection{Classification result} \label{secsl}
We report the classification performance of all methods in Figure \ref{fig1}. We also report the classification performance of our proposed method using single-task learning and multi-task learning  in Table \ref{tab2} and the Receiver Operating Characteristic (ROC) curves of all methods in Figure \ref{fig1}. Based on the results, we  conclude our observations as follows.

First, it could be observed that the proposed method achieves the best classification performance, followed by SFS, MSFS, random forest, L1SVM, and SVM. Specifically, our proposed method improve on average by 32.80\% and 11.13\%, respectively, compared to the worst comparison method (\ie SVM) and the best comparison methods (\ie SFS), in terms of all four evaluation metrics. The reason is the facts that our method takes into account the issues in the same framework, such as feature selection removing the redundant features, sample weight reducing the influence of the outliers and solving the problem of imbalance classification to reduce the issue of high false negatives, and joint classification and regression utilizing the share information between two tasks to improve the model effectiveness of each of them.

Second, it is important to conduct feature selection for analyzing  high-dimensional data. High-dimensional data easily results in the issue of curse of  dimensionality. In the literature, many studies showed that the classification model on the high-dimensional data will output low performance \cite{zhu2014novel,adeli2018semi}.  Figure \ref{fig1} verified the above statement.  In our experiments, only SVM does not consider the issue of high-dimensional data and achieves the worst classification performance. More specifically, the  best comparison method (\ie SFS) improves by 21.09\%  and the worst  comparison method (\ie L1SVM) improve by 4.73\%, for all evaluation metrics, compared to the baseline SVM.

Third, it is useful to use joint classification and regression framework for detecting the severe cases from mild confirmation cases. As shown in Table \ref{tab2} and Figure \ref{fig1}, our proposed method conducting joint regression and classification achieves better classification performance, compared to the single-task based classification methods, \eg random forest, L1SVM, SFS, and SVM. Moreover, both MSFS and our method are joint models. However, our method outperforms MSFS since our proposed method takes into account one more constraint, \ie imbalance classification. In particular, we conducted single-task classification using Eq. (\ref{eq3}), \ie our proposed method without considering the regression task, Proposed w/o Regression in Table \ref{tab2}. As a result, Proposed w/o Regression outperforms SFS since both of them take into account three following constraints, such as feature selection, sample weight, and imbalance classification.


\begin{table}[!tb]
	\centering
	\caption{Regression results  of all methods.}
	\centering
	{\scriptsize
		\begin{tabular}[c]{|c|c|c|}\hline
			Methods   &CC  &RMSE  \\\hline			
Ridge regression                  &0.329 $\pm$ 0.158  &20.02 $\pm$ 9.724   \\ \hline
L1SVR  \cite{chang2011libsvm}     &0.351 $\pm$ 0.085  &10.49 $\pm$ 2.072  \\ \hline
Lasso \cite{tibshirani1996regression}  &0.354 $\pm$ 0.165  &9.92 $\pm$ 9.571 \\ \hline
Random forest \cite{liaw2002classification} &0.406 $\pm$ 0.188  &13.22 $\pm$ 6.762   \\ \hline
MSFS    \cite{zhu2014novel}   &0.408 $\pm$ 0.092  &9.29 $\pm$ 1.104    \\ \hline
Proposed    &\textbf{0.462} $\pm$ 0.056  &\textbf{7.35}  $\pm$ 1.087     \\ \hline
		\end{tabular}		
	}
\label{tab3}
\end{table}

\subsection{Regression results} \label{secsl}

We evaluated the regression performance through the prediction of conversion time from the non-severe case to the severe case. We report the results of correlation coefficients (CCs) and RMSEs of all  methods in Table \ref{tab3}.

First, the regression performance of the methods without feature selection (\eg ridge regression) is worse  than methods with feature selection, \eg Lasso, L1SVR, MSFS, and ours. Moreover, our method outperforms all comparison methods. For example, our method receives the best performance for correlation coefficient (\eg 0.462) and RMSE (\eg 7.351).

Second, similar to the results of the classification task, the results of the regression task  show the advantages of the considerations, such as feature selection, sample weight, imbalance classification, and joint classification and regression.  In particular, our proposed method considering all four considerations improves 0.054 and 1.940, respectively, in terms of correlation coefficient and RMSE, compared to  MSFS which takes two considerations into account, such as feature selection, and joint classification and regression.

\begin{table}[!tb]
	\centering
	\caption{Regions distribution at different HU ranges for top selected regions.}
	\centering
	{\scriptsize
		\begin{tabular}[c]{|c|c|c|}\hline
Hu ranges           & left lung (6) & right lung (16)\\\hline
$[-\infty, -700]$   &0  &2 \\ \hline
$[-700, -500]$      &1  &8 \\ \hline
$[-500, -200]$      &2  &5 \\ \hline
$[-200, 50]$        &1  &0 \\ \hline
$[50, \infty]$      &2  &1 \\ \hline
\end{tabular}		
	}
\label{tab4}
\end{table}

\begin{figure*}[!t]
 \begin{center}
  \subfigure[Ridge regression] {\scalebox{0.315}
   {\includegraphics{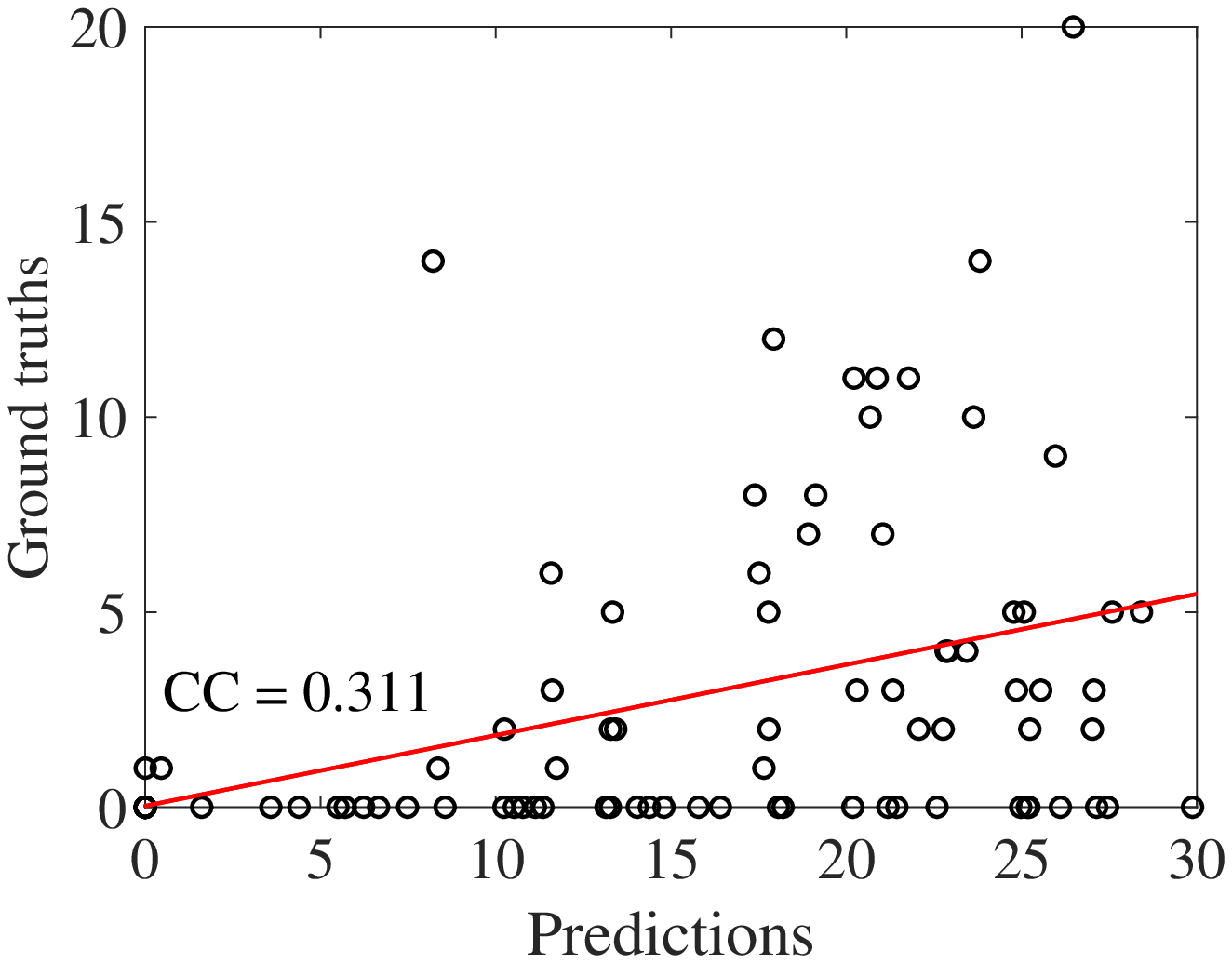}}}
  \subfigure[L1SVR]      {\scalebox{0.315}
   {\includegraphics{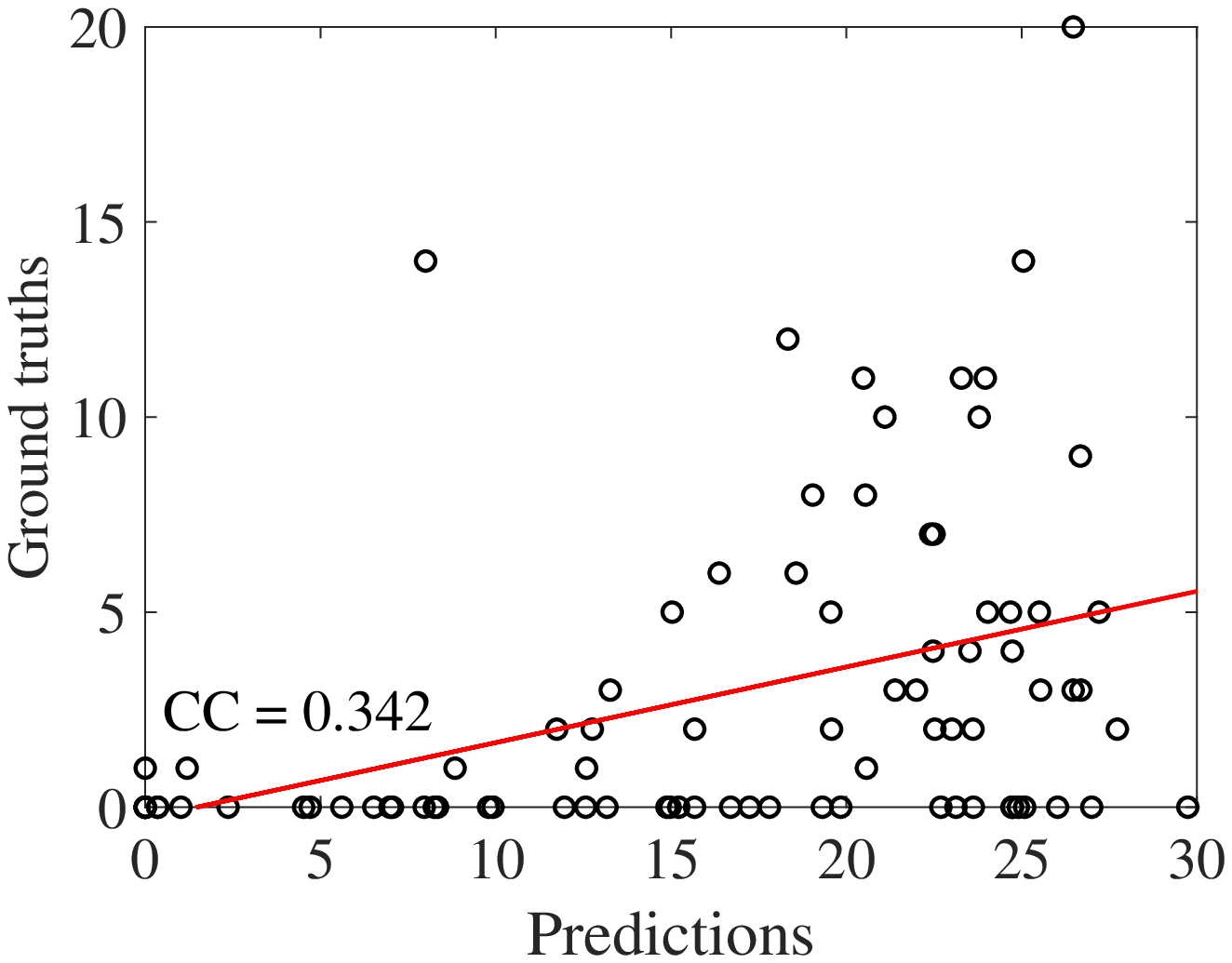}}}
  \subfigure[Lasso]      {\scalebox{0.315}
   {\includegraphics{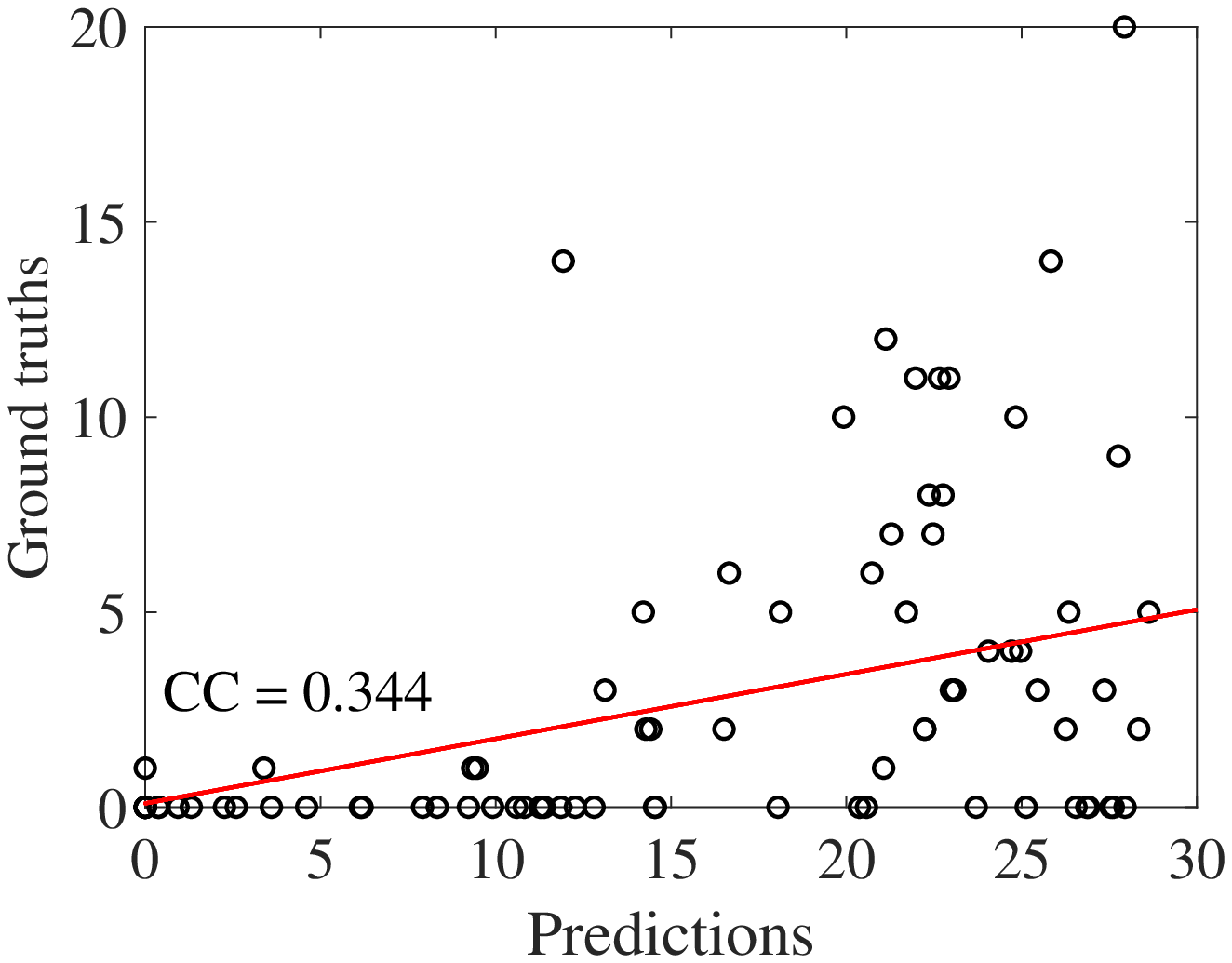}}}
  \subfigure[Random forest]    {\scalebox{0.315}
   {\includegraphics{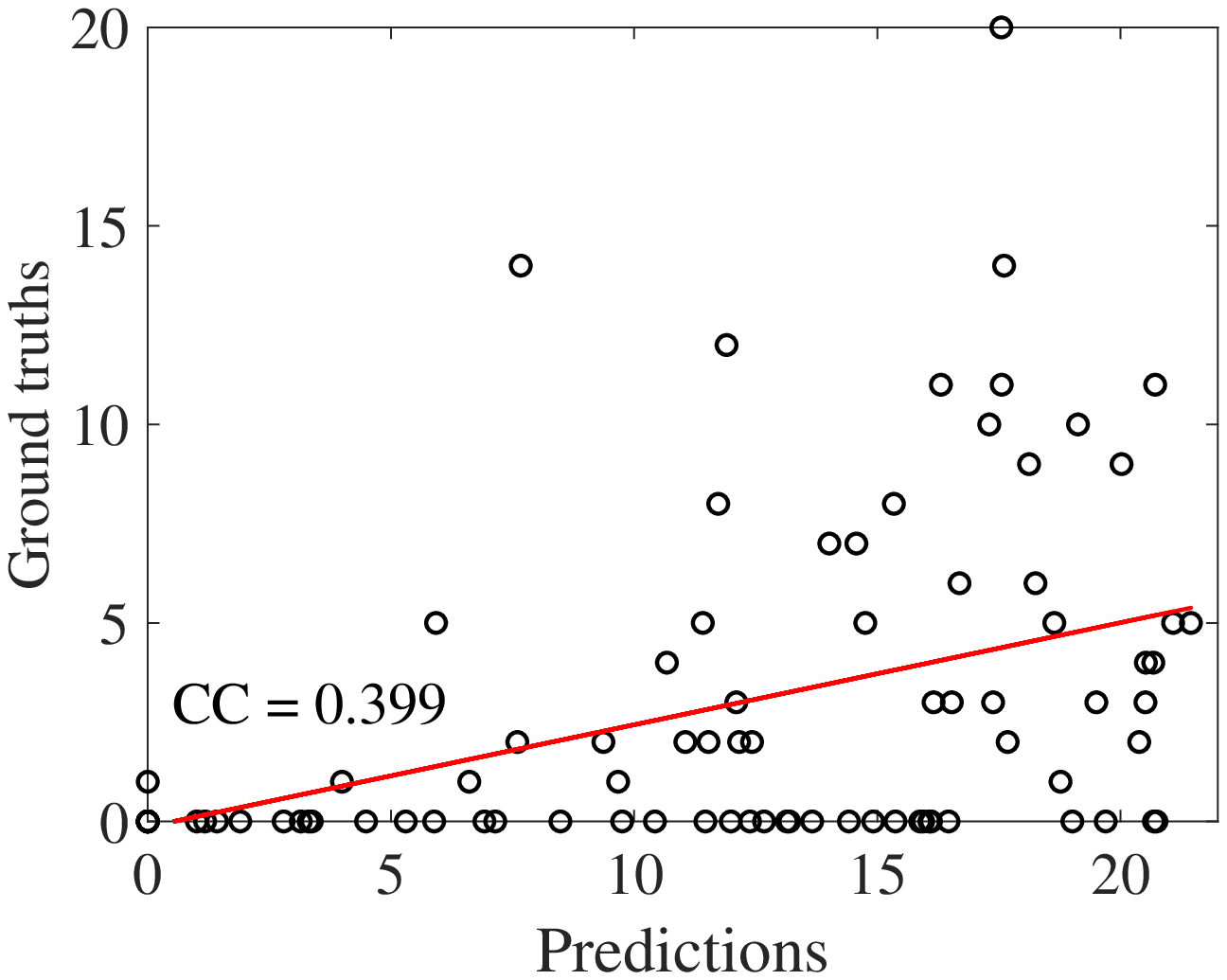}}}
  \subfigure[MSFS]    {\scalebox{0.315}
   {\includegraphics{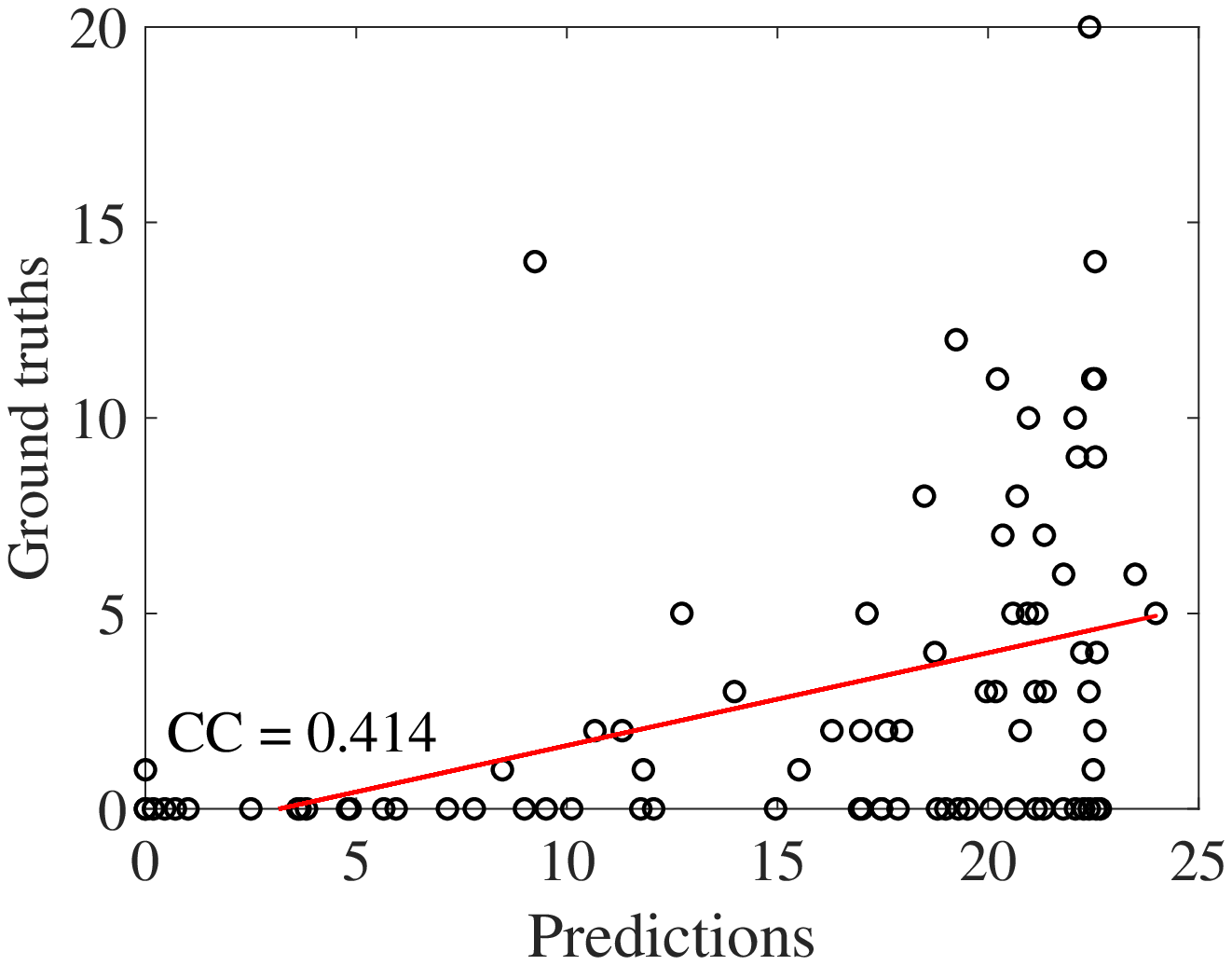}}}
  \subfigure[Proposed] {\scalebox{0.315}
   {\includegraphics{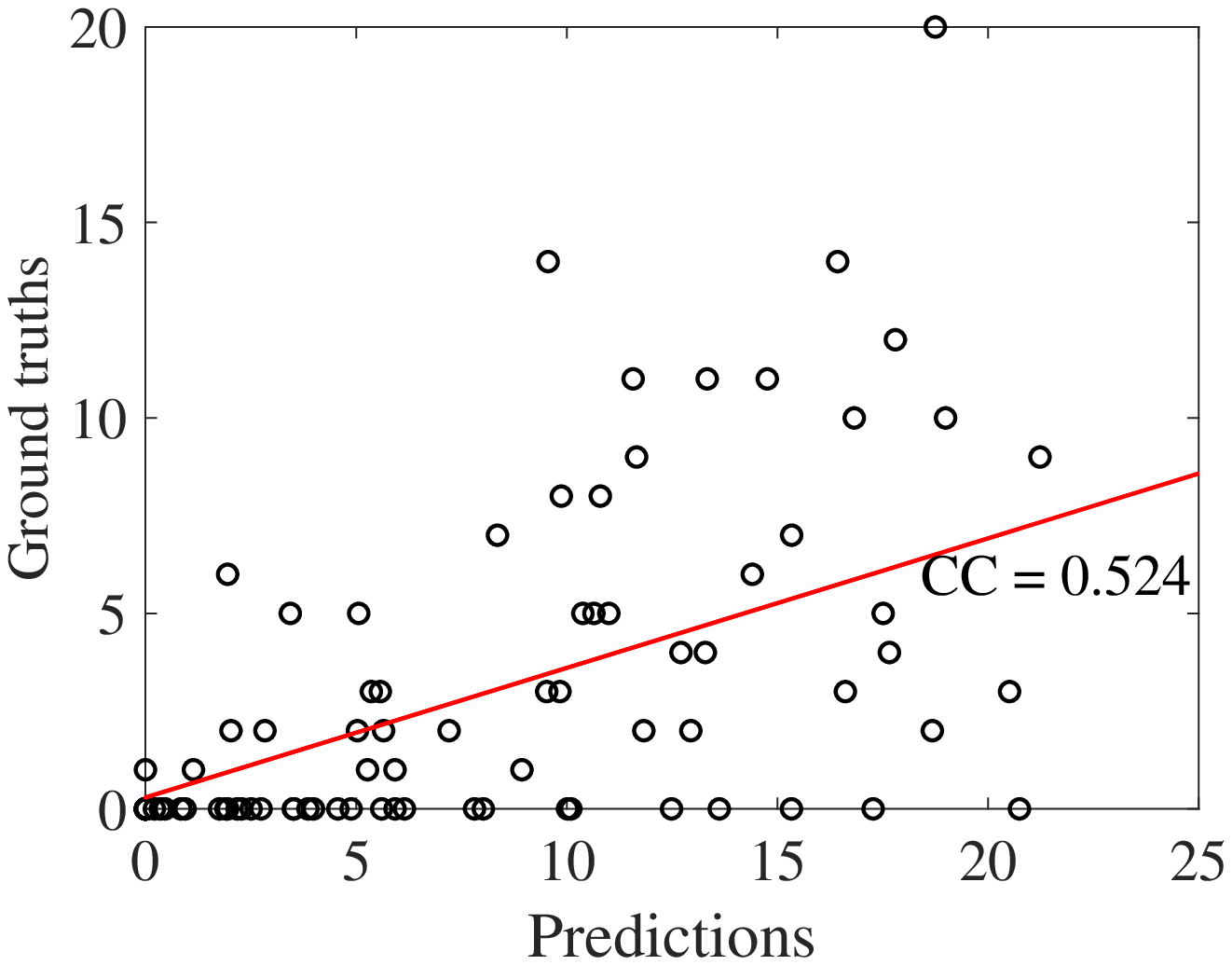}}}
   \vspace{-3mm}
  \caption{Scatter plots and the corresponding correlation coefficients (CCs) of all methods for predicting the severe cases.}
  \label{fig3}
 \end{center}
\end{figure*}

\begin{figure*}[!t]
 \begin{center}
  \subfigure {\scalebox{1} {\includegraphics{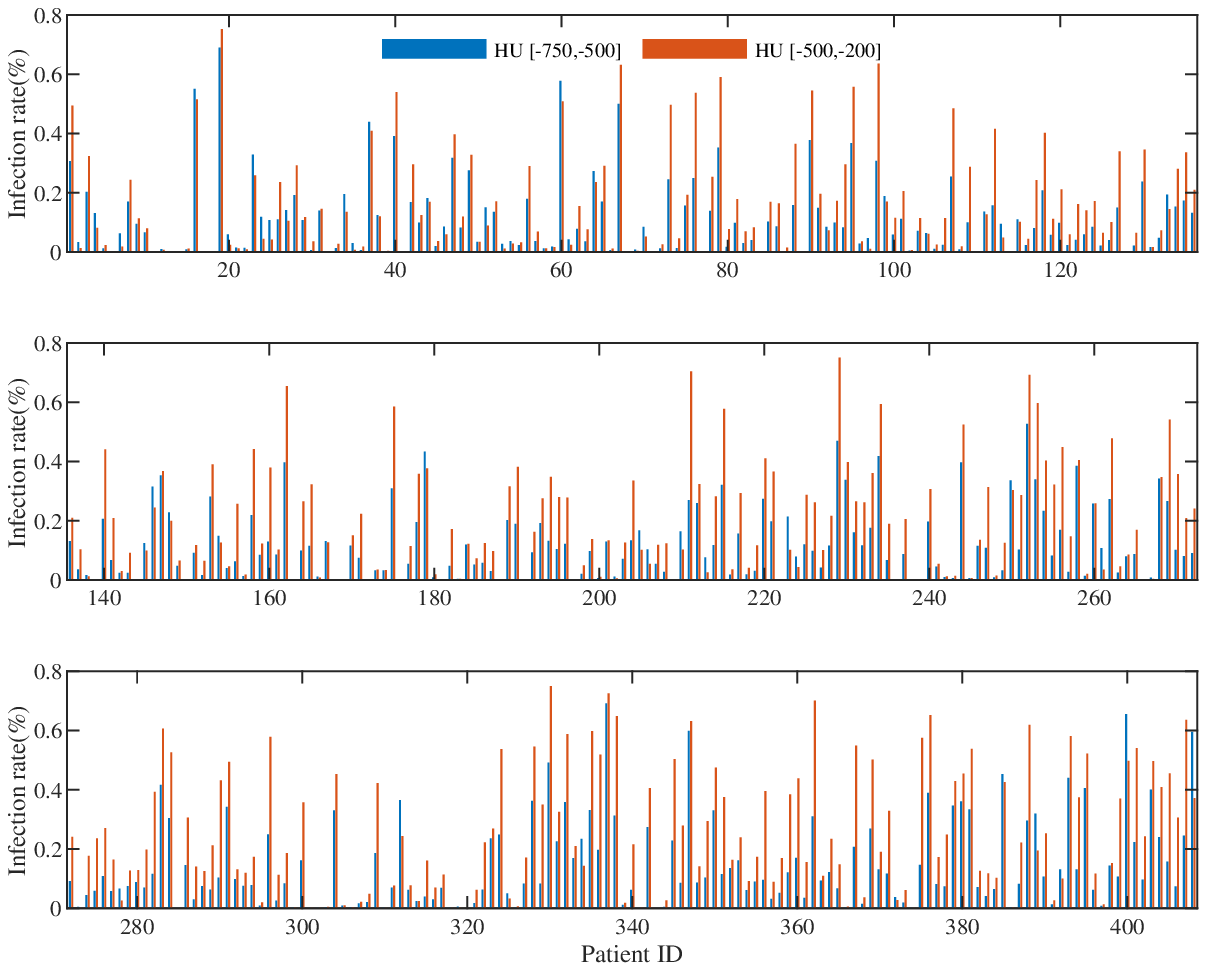}}}
  \vspace{-3mm}
  \caption{\footnotesize Ratios of infected volumes in the HU ranges of [-700,-500] and [-500, -200], where patient IDs 1-322  are non-severe cases and patient IDs (323-408) are severe cases.}
  \label{fig4}
 \end{center}
\end{figure*}

\section{Discussion}
\subsection{Imbalance classification}
In the classification task, our method investigates the issues, \ie feature selection, sample weight, imbalance classification, and joint classification and regression. As a result, our method outperforms all comparison methods only focusing on part of four issues. Moreover, our solution for each issue is shown reasonable and feasible. An interesting question is which issue dominates the  COVID-19 analysis with chest CT scan data. There is not theoretical answer. However, based on our experimental results, the problem of imbalance classification should be the first issue to be considered due to the following reasons.

First, it is necessary to take into account the problem of imbalance classification. In our experiments, random forest outperforms L1SVM (\eg 6.84\% for all evaluation metrics) because random forest considers the problem of imbalance classification and L1SVM takes into account the issue of high-dimensional data. Moreover, the only difference between  SFS and MSFS is that SFS considers the problem of imbalance classification and MSFS conducts a joint classification and regression. As a result, SFS beats MSFS a little bit, \ie 1.26\% improvement in terms of all evaluation metrics.

Second, in Figure \ref{fig1}, the sensitivities of the methods (\eg SVM, L1SVM, and MSFS) are low, \eg 23.86\%, 26.73\%, and 47.45\%  respectively. The reason is that their classifiers could directly predict the samples of the minority class with the label of the majority class to output high accuracy,  \eg 67.35\%, 75.26\%, and 86.31\%, respectively. On the contrary, the methods (\eg random forest,  SFS, and our Proposed w/o Regression) consider the issue of imbalance classification to output the high sensitivities, \eg 49.61\%, 50.65\%, and 70.73\%, respectively.

\subsection{Top selected regions}
In this paper, we did not employ deep learning methods due to the interpretability and the issue of small-sized sample. In this section, we list  top selected features (\ie the chest regions) in Table \ref{tab4}, which could help the clinicians to improve the efficiency and the effectiveness of the disease diagnosis.
To do this,   we first obtained the totally  selected number for each feature across 100 experiments, \ie repeating the 5-fold cross validation scheme 20 times, and then reported top 22 selected features (\ie regions), each of which was selected at least 90 out of 100 times. We list our observations as follows.

First, most of selected features (\ie 17 out of 22) are in the HU range of $[-700,-200]$, corresponding to the regions of ground glass opacity which has been demonstrated related to the severity of COVID-19  \cite{tang2020severity}. Second,  the region number in the right lung is larger than  the number in the left lung, \ie 16 vs. 6. The possible reason is that the virus might easily  infect  the regions in the right lung \cite{shi2020radiological}. Third, we extracted 3 kinds of handcrafted features, \ie density, mass, and volume, from each part. Moreover, the mass feature is related to both the density feature and the volume feature. Based on the results, our method selected 4 and 7 density features, respectively, from the left lung and the right lung, and 2 and 6 mass features, respectively, from the  left lung and the right lung. However, our method only selected 3 volume feature from the right lung. Hence, we would have the conclusion that the density feature is the most important in our experiments, followed by the mass feature and the volume feature.

\subsection{Importance of prediction and time estimation of severe cases}
To our knowledge, this study is the first work to simultaneously predict and estimate the conversion time of COVID-19 developing to severe symptoms using chest CT scan data.

First, our method obtains higher sensitivity, \ie 76.97\%, compared to \cite{tang2020severity}, \ie 74.5\% of sensitivity. That is, our method achieves higher accuracy for classifying the severe cases than \cite{tang2020severity}. The reason could contribute to that 1) our  method designs a novel solution for the problem of imbalance classification, and 2) the regression information in our proposed joint  model improves the classification performance.

Second, as shown in Figure \ref{fig3}, the correlation coefficient (\ie 0.524)  between our predictions and the corresponding ground truths for the severe cases is larger than the value (\ie 0.462) in Table \ref{tab3} which measures the correlation between our predictions and the corresponding ground truths for all subjects. Moreover, our proposed method yields the averaged conversion time (\ie 4.59 $\pm$ 0.223 days, which has 0.55 days different from the ground truth of the conversion time, \ie $5.64 \pm 4.30$ days) from all non-serve cases to the severe stage, with the least estimation error (\ie 6.01 $\pm$ 1.22), compared to all comparison methods. The possible reason should be the proposed  joint classification and regression model, where the classification information could improve the effectiveness of the regression task. Above advantages of our proposed method imply that our proposed method is good at predicting the conversion time from the non-severe stage to the severe stage.

Above two observations indicate that our proposed method is suitable for predicting the severe cases. In real applications, correctly classifying severe cases is more important than correctly classifying the non-severe cases because  the former could  reduce the clinicians�workloads. In particular, the correct prediction of the conversion time could help the clinicians designing effective treatment plan for the potential severe cases in time or even save the patients' lives.


\subsection{Limitations}
This study yielded an accuracy of 85.91\%, which seems lower than that reported in previous severity assessment work. However, the task in this paper tried to solve is quite different, as we predict whether the patient would develop severe symptom in the later time. This would result in the problem of imbalance classification since only a small portion of patients would convert severe based on the prevalence rate.
First, the problem of imbalance classification of our data set  is bias, \ie 86 severe cases vs. 322 non-severe cases. This makes difficult to construct effective classification models. Second,  the difference of infected volumes between the severe cases and the non-severe cases is small, as shown in Figure \ref{fig4}, while the corresponding difference is distinguished  in \cite{tang2020severity}, thus the latter can  easily conduct classification.
With the increase of available the data of severe cases, the accuracy of our method could be further improved.
In our future work, we plan to generate new samples for the minority class to lessen the problem of imbalance classification, as well as design new deep transfer learning methods using other data sources (\eg X-ray data) to solve the issue of small-sized sample and high-dimensional features.

This study only focused on binary classification, \ie severe cases vs. non-severe cases. In our future work, we plan to conduct multi-class classification on four types of COVID-19 diagnosis, \ie mild,  common,  severe,  and  critical.

\section{Conclusion}
In this paper, we proposed a new method to jointly conduct disease identification and conversion time prediction, by taking into account the issues, such as high-dimensional data, small-sized sample, outlier influence, and imbalance classification. To do this, we designed a sparsity regularization term to conduct  feature selection and learn the shared information between two tasks, and proposed a new method to take into account the sample weight and the issue of imbalance classification.
Finally, experimental results showed that our proposed method achieved the best performance for detecting the severe case from  non-severe cases and the conversion time from the mild confirmed case to the severe case with the CT data in a real data set, compared to the comparison methods.

\small{
\bibliographystyle{elsarticle-harv}
\bibliography{covid}
}

\end{document}